\documentclass[aps,preprint,preprintnumbers,amsmath,showpacs,amssymb,nofootinbib]{revtex4}
\usepackage{amssymb}
\usepackage{amsmath}
\usepackage{bm}
\usepackage{graphicx}
\draft
\begin{document}
\title{Dirac equation for massive neutrinos in a Schwarzschild-de Sitter spacetime from a 5D vacuum.}
\author{$^{1}$ Pablo Alejandro S\'anchez\footnote{pabsan@mdp.edu.ar}, Mariano Anabitarte\footnote{anabitar@mdp.edu.ar} and Mauricio
Bellini\footnote{mbellini@mdp.edu.ar}}
\address{$^{1}$ Departamento de F\'{\i}sica, Facultad de Ciencias Exactas y
Naturales, \\
Universidad Nacional de Mar del Plata, Funes 3350, (7600) Mar del
Plata, Argentina. \\ \\
$^{2}$ Instituto de Investigaciones F\'{\i}sicas de Mar del Plata
(IFIMAR), Consejo Nacional de Investigaciones Cient\'{\i}ficas y
T\'ecnicas (CONICET), Argentina.}
\begin{abstract}
Starting from a Dirac equation for massless neutrino in a 5D
Ricci-flat background metric, we obtain the effective 4D equation
for massive neutrino in a Schwarzschild-de Sitter (SdS) background
metric from an extended SdS 5D Ricci-flat metric. We use the fact
that the spin connection is defined to an accuracy of a vector, so
that the covariant derivative of the spinor field is strongly
dependent of the background geometry. We show that the mass of the
neutrino can be induced from the extra space-like dimension.
\end{abstract}
\maketitle

\section{Introduction}

The study of wave scattering in black hole spacetimes is crucial
to the understanding of the signals expected to be received by the
new generation of gravitational-wave detectors in the near
future\cite{..}. Since the linear perturbations of black holes are
represented by fields of integral spins, the study of the
scattering of wave fields are concentrated on these cases while
that of the Dirac fields are thus less common, especially for the
massive ones\cite{...}. The neutrino does not respond directly to
electric or magnetic fields. Therefore, if one wishes to influence
its orbit by forces subject to simple analysis, one has to make
use of gravitational fields. In other words, one has to consider
the physics of a neutrino in a curved metric. There are both
numerical and analytical methods in solving the various wave
equations in black hole scattering\cite{1}. Other approximated
solutions can be obtained using the semi-analytic WKB
approximation\cite{2}, which has been proven to be very useful and
accurate in many cases like, for example, the evaluation of the
quasi normal mode frequencies\cite{3}. Quasi normal modes
frequencies has been obtained using approximated analytical
solutions with SUSY for massless neutrinos\cite{3b}.

On the other hand, the extension of 4D spacetime to $N(\geq 5)D$
manifolds is the preferred route to a unification of the
interactions of particle physics with gravity. The two approaches
to 5D relativity of most current interest are brane theory and
induced matter theory, which are mathematically
equivalent\cite{w2}. The former commonly uses the warp metric in
5D, and in general leads to an extra force on a massive particle
in 4D\cite{w3}, where however the particle can travel on a null
path in 5D\cite{w4}. Similar results were found previously for
induced-matter theory, which commonly uses the canonical metric in
5D to isolate the fifth force\cite{w5} and examine null geodesics
in 5D\cite{w6}. These investigations are classical in nature, but
clearly invite an examination of the corresponding picture in
quantum theory. In a more recent work was examined the 4D
Klein-Gordon and Dirac equation from 5D null paths\cite{wes}.

In this letter we shall study the 4D Dirac equation for neutral
$1/2$-spin fermions with mass (neutrinos) which are close to a
non-rotating  SdS black hole, but using a extended SdS, which
describes a 5D vacuum on a 5D Ricci-flat metric. We shall consider
that the space-like extra coordinate is noncompact.

\section{The Dirac equation for massless neutrinos on a 5D vacuum}

We shall consider the 5D Ricci-flat ($R_{AB}=0$ and then the Ricci
scalar $R=0$) metric $g_{AB}$, given by the line element\cite{mb}
\begin{equation}\label{a1}
dS^{2}=\left(\frac{\psi}{\psi_0} \right)^{2}\left[-
c^{2}f(r)dt^{2}+ \frac{dr^2}{f(r)}+r^{2}(d\theta ^{2}
+sin^{2}\theta d\phi ^{2})\right]+d\psi^{2},
\end{equation}
where
$f(r)=1-(2G\zeta\psi_{0}/(rc^2))[1+c^{2}r^{3}/(2G\zeta\psi_{0}^{3})]$
is a dimensionless function, $\lbrace t,r,\theta,\phi\rbrace$ are
the usual local spacetime spherical coordinates employed in
general relativity and $\psi$ is the noncompact space-like extra
dimension. Furthermore, $\psi$ and $r$ have length units, $\theta$
and $\phi$ are angular coordinates, $t$ is a time-like coordinate,
$c$ denotes the speed of light, $\psi_0$ is an arbitrary constant
with length units and the constant parameter $\zeta$ has units of
$(mass)(length)^{-1}$.

We are concerned with a 5D spacetime as a main bundle and $U(1)$
its structural group. Einstein's 4D spacetime is the usual base of
this bundle, so, physical fields will not depend on the extra
coordinate.

\subsection{The 5D Clifford algebra for spinors}

To define a 5D vacuum we shall consider a Lagrangian for a
massless 5D spinor field minimally coupled to gravity
\begin{equation}
L = \frac{\hbar c}{2} \left[ \bar{\Psi} \gamma^A \left(\nabla_A
\Psi\right) - \left( \nabla_A \bar{\Psi}\right)
\gamma^A\Psi\right] + \frac{R}{2K},
\end{equation}
where $K={8\pi G\over c^4}$ and $\gamma^A$ are the Dirac matrices
which satisfy
\begin{equation}
\left\{ \gamma^A, \gamma^B\right\} = 2 g^{AB} \;\mathbb{I},
\end{equation}
such that the covariant derivative of the spinor $\Psi$ on
(\ref{a1}) is defined in the following form:
\begin{equation}
\nabla_A \Psi = \left( \partial_A + \frac{1}{8} \Gamma_A \right)
\Psi,
\end{equation}
and the spin connection is given by
\begin{equation}
\Gamma_A =\frac{1}{8} \left[ \gamma^b, \gamma^c\right] e_b^B\,
\nabla_A\left[ e_{cB}\right],
\end{equation}
$\nabla_A\left[ e_{cB}\right]=\partial_A e_{cB} - \Gamma^D_{AB}
e_{cD}$ being the covariant derivative of the five-bein $e_B^c$
(the symbol $\partial_A$ denotes the partial derivative with
respect to $x^A$ and $\eta_{ab}= g_{AB} e^A_{\,\,a} e^B_{\,\,b}$
denotes the 5D Minskowsky spacetime in cartesian coordinates),
which we introduce in order to generalize the well known 4D
vierbein\cite{chotin}, but to relate the extended Schwarzschild-de
Sitter metric (\ref{a1}) with the 5D Minkowsky spacetime written
in cartesian coordinates: $dS^2= -c^2 dt^2 +
(dx^2+dy^2+dz^2)+d\psi^2$
\begin{equation}
e_B^c= \left( \begin{array}{lllll} \left(\frac{\psi}{\psi_0}\right) c \sqrt{f(r)} & 0 & 0 & 0 & 0 \\
0 & \left(\frac{\psi}{\psi_0}\right) \frac{\sin{\theta}
\cos{\phi}}{\sqrt{f(r)}} & \left(\frac{\psi}{\psi_0}\right)
\frac{\sin\theta \sin\phi}{\sqrt{f(r)}} &
\left(\frac{\psi}{\psi_0}\right) \frac{\cos\theta}{\sqrt{f(r)}}
&  0 \\
0 & \left(\frac{\psi}{\psi_0}\right) r\,\cos\theta \cos\phi&
\left(\frac{\psi}{\psi_0}\right) r\,\cos\theta \sin\phi & -
\left(\frac{\psi}{\psi_0}\right) r\,\sin\theta
& 0 \\
0 & -\left(\frac{\psi}{\psi_0}\right) r\,\sin\theta \sin\phi &
\left(\frac{\psi}{\psi_0}\right) r\,\sin\theta
\cos\phi & 0 & 0 \\
0 & 0 & 0 & 0 &  1\end{array} \right).
\end{equation}
The Dirac matrices $\gamma^a$ are represented in an Euclidean
space instead in a Lorentzian space, and are described by the
algebra\cite{grov,camporesi}: $\left\{ \gamma^a,\gamma^b\right\} =
2 \eta^{ab} \mathbb{I}$
\begin{eqnarray}
&& \gamma^0= \left( \begin{array}{llll} -i & 0 & 0 & 0  \\
0 & -i & 0 &  0 \\
0 & 0 & i & 0 \\
0 & 0 & 0 & i
\end{array} \right) = \left(\begin{array}{ll} -i \mathbb{I} & 0 \\
0 & i \mathbb{I}\ \end{array} \right),\qquad
\gamma^1= \left( \begin{array}{llll} 0 & 0 & 0 & -i \\
0 & 0 & -i & 0 \\
0 & i & 0 & 0 \\
i & 0 & 0 & 0 \end{array} \right) = \left(\begin{array}{ll} 0 & -i \sigma^1 \\
i \sigma^1 & 0  \end{array} \right),  \nonumber \\
&& \gamma^2= \left( \begin{array}{llll} 0 & 0 & 0 & -1 \\
0 & 0 & 1 & 0 \\
0 & 1 & 0 & 0 \\
-1 & 0 & 0 & 0 \end{array} \right) = \left(\begin{array}{ll} 0 & -i \sigma^2 \\
i \sigma^2 & 0  \end{array} \right),  \qquad \gamma^3= \left( \begin{array}{llll} 0 & 0 & -i & 0 \\
0 & 0 & 0 & i \\
i & 0 & 0 & 0 \\
0 & -i & 0 & 0 \end{array} \right) = \left(\begin{array}{ll} 0 & -i \sigma^3 \\
i \sigma^3 & 0  \end{array} \right),\nonumber \\
\end{eqnarray}
such that $\gamma^4 = i \gamma^0 \gamma^1 \gamma^2 \gamma^3$, and
the $\sigma^i$
\begin{eqnarray}
&& \sigma^1 = \left(\begin{array}{ll} 0 & 1 \\
1  & 0  \end{array} \right), \qquad \sigma^2 = \left(\begin{array}{ll} 0 & -i \\
i  & 0  \end{array} \right), \qquad \sigma^3 = \left(\begin{array}{ll} 1 & 0 \\
0  & -1  \end{array} \right),
\end{eqnarray}
are the Pauli matrices.

The stress tensor for free massless neutrinos is 
\begin{equation}
T^a_{\,\,b} = -\left\{\frac{1}{4} \left[ \bar\Psi,\gamma^a
\nabla_b \Psi\right]+ \frac{1}{4} \left[ \Psi,\gamma^a \nabla_b
\bar\Psi\right]\right\},
\end{equation}
For a Ricci-flat metric, like (\ref{a1}), the expectation value of
this tensor would be zero: $\left< T^a_{\,\,b} \right> =0$.
Therefore, in a Ricci-flat metric the following condition must be
hold
\begin{equation}
 \left[ \bar\Psi,\gamma^a \nabla_b \Psi\right] =
\left[\gamma^a \nabla_b \bar\Psi, \Psi\right].
\end{equation}

\subsection{The Dirac equation for spinors in 5D}

Finally, using the fact that $ \gamma^A = e^A_a \gamma^a$, we
obtain that the Dirac equation on the metric (\ref{a1}), when we
use 3D spherical cordinates $(r,\theta,\phi)$, is
\begin{eqnarray}
\gamma^0 \frac{1}{c \sqrt{f(r)}} \frac{\partial \Psi}{\partial t}
&+& \gamma^r \frac{[f(r)]^{1/4}}{r} \frac{\partial}{\partial r}
\left[ r [f(r)]^{1/4} \Psi\right] - \frac{\gamma^r}{r} \left(
\vec{\Sigma} . \vec{L} + \mathbb{I}\right) \Psi \nonumber \\
&+& \gamma^4 \left[ \left(\frac{\psi}{\psi_0}\right)
\frac{\partial\Psi}{\partial\psi}+ \frac{2}{\psi_0} \Psi\right]
=0,  \label{dirac}
\end{eqnarray}
where $\gamma^r$ is defined as
\begin{equation}
\gamma^r=\gamma^{1}{\rm sin}\theta\ \!{\rm cos}\phi\
\!+\gamma^{2}{\rm sin}\theta\ \!{\rm sin}\phi+\gamma^{3}{\rm
cos}\theta,
\end{equation}
and the ordinary angular momentum operators are
\begin{equation}
\vec{\Sigma}=\left(
\begin{array}{cc}
\vec{\sigma}&0\\0&\vec{\sigma}
\end{array}\right), \qquad \vec{L}=\vec{r}\times\vec{p},
\end{equation}
so that
\begin{eqnarray}
\gamma^r \left(\vec{\Sigma} . \vec{L}\right) &= & \gamma^1 \left(
- \cos{\theta} \cos{\phi} \frac{\partial}{\partial\phi}+
\frac{\sin{\phi}}{\sin{\theta}}
\frac{\partial}{\partial\phi} \right) \nonumber \\
&+& \gamma^2 \left(-\cos{\theta}\sin{\phi}
\frac{\partial}{\partial\theta} - \frac{\cos{\phi}}{\sin{\theta}}
\frac{\partial}{\partial\theta}\right) +\gamma^3 \sin{\theta}
\frac{\partial}{\partial\theta}.
\end{eqnarray}

We consider the 5D Dirac equation (\ref{dirac}). We can make the
ansatz: $ \Psi(t,r,\theta\phi,\psi) = \bar{\Psi}(t,r,\theta,\phi)
\, \Upsilon(\psi)$, such that the equation for $\Upsilon(\psi)$ is
\begin{equation}\label{sep}
\left[ \left(\frac{\psi}{\psi_0}\right)
\frac{\partial\Upsilon(\psi)}{\partial\psi}+
\frac{2}{\psi_0}\Upsilon(\psi) \right]
 = m\, \Upsilon(\psi),
\end{equation}
$m=m_0/\psi_0$ being a separation constant.

\subsection{The mass of neutrinos in a 5D vacuum}

\begin{figure*}
\includegraphics[height=15cm]{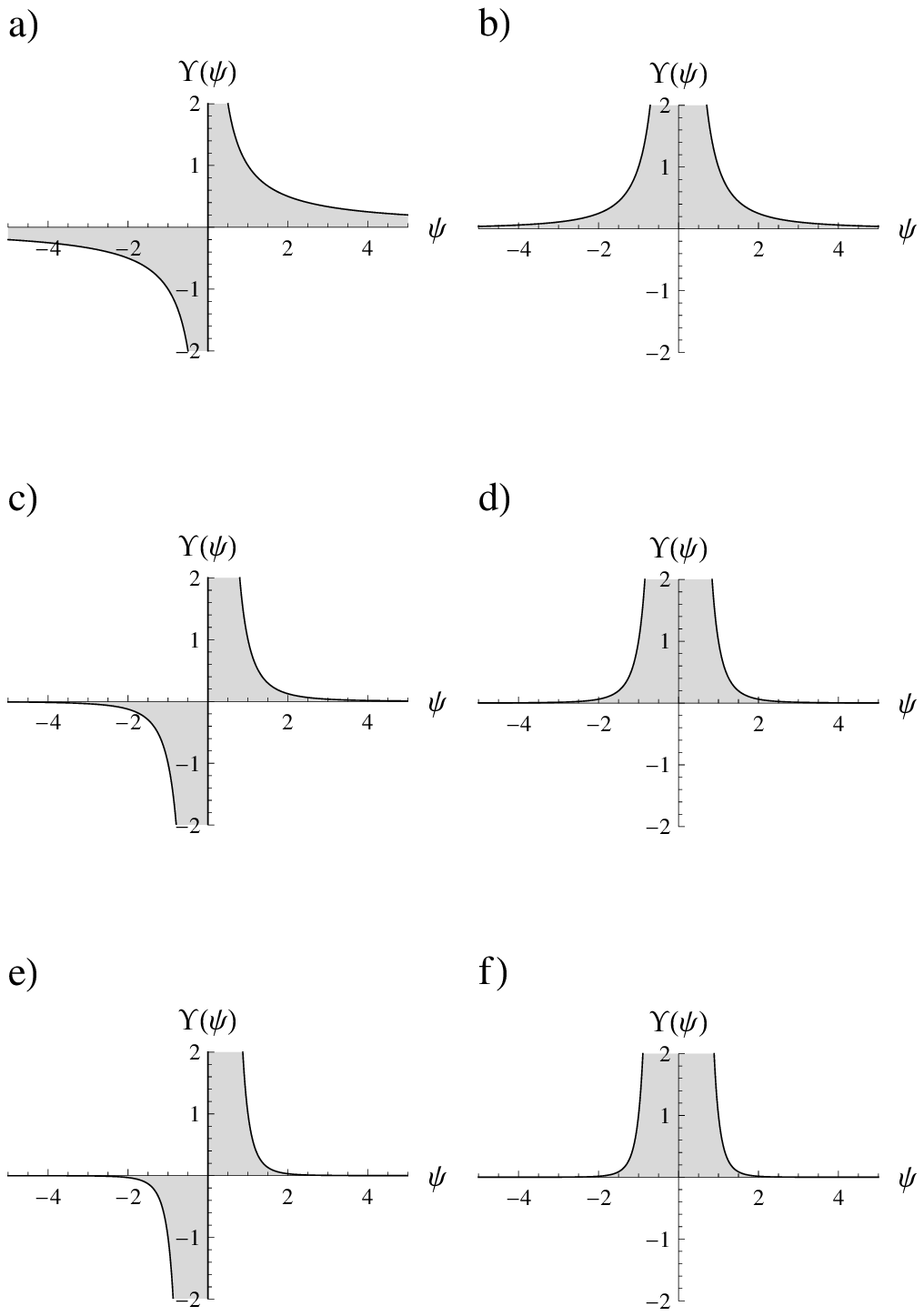}\caption{\label{figura1} The function $\Upsilon(\psi)$ for diferent values of
$m_0$: a) $m_0 =1$, b) $m_0 = 0$, c) $m_0 = -1$, d) $m_0 = -2$, e)
$m_0 = -3$ and f) $m_0= -4$.}
\end{figure*}

The solution for the equation (\ref{sep}) is
\begin{equation}
\Upsilon(\psi) =
\Upsilon_0\left(\frac{\psi}{\psi_0}\right)^{m_0-2},
\end{equation}
where $\Upsilon_0$ is a constant of integration. On the other
hand, for $m_0 < 2$ the function $\Upsilon(\psi)$ tends to $0$ for
$\psi \rightarrow \pm \infty$, but is divergent for $\psi
\rightarrow 0$. In order to the function $\Upsilon(\psi)$ to be
real, we must ask $m_0$ take integer unbounded values: $m_0 = ...
,2,1,0,-1,-2, ...$. It appears to be a form of quantization. In
the figure (\ref{figura1}) we have plotted $\Upsilon(\psi)$ for
different values of $m_0$. Notice that for even $|m_0|$ values the
function $\Upsilon(\psi)$ is even but for odd $|m_0|$ values the
function is also odd. Finally, one could distinguish between two
different cases. {\bf i)} For ($\psi_0 >0$ and $m_0 \geq 0$) or
($\psi_0 <0$ and $m_0 \leq 0$), one obtains $m >0$. {\bf ii)} For
($\psi_0 >0$ and $m_0 \leq 0$) or ($\psi_0 <0$ and $m_0 \geq 0$),
the mass is negative: $m <0$. The last case is a nonsense result
in 4D physics and therefore should be discarded.

\section{The induced 4D Dirac equation for massive neutrinos close to a SdS spacetime}

Now let us to assume that the 5D spacetime can be foliated by the
family of hypersurfaces $\lbrace\Sigma _{0} :\psi=\psi
_{0}\rbrace$. On every generic hypersurface $\Sigma_{0}$  the
induced metric is given by the 4D line element
\begin{equation}\label{a2}
dS^{2}_{ind}=-c^{2}f(r)dt^{2}+\frac{dr^{2}}{f(r)}+r^{2}(d\theta
^{2}+sin^{2}\theta d\phi ^{2}),
\end{equation}
which describes a Schwarzschild-de Sitter spacetime with an
equation of state $\omega = P/(c^2 \rho)= -1$. If we assume a
static foliation of the 5D spacetime on the 4D hypersurface
$\Sigma_0$, the 4D energy momentum tensor will be described by a
perfect fluid $\bar{T}_{\alpha\beta}= e^{A}_{\alpha} e^B_{\beta}
T_{AB} =(\rho c^2 +P)u_{\alpha}u_{\beta}-P h_{\alpha\beta}$, where
$\rho(t,r)$ and $P(t,r)$ are respectively the energy density and
pressure of the induced matter. Furthermore, the 4-velocities
$u_{\alpha}$ are related to the 5-velocities $U_A$ by $u_a =
e^A_{\alpha} U_A$, and $h_{\alpha\beta} = e^A_{\alpha} e^B_{\beta}
g_{AB}$ are the components of the tensor metric in (\ref{a2}).
From the relativistic point of view, observers that are on
$\Sigma_0$ move with $U^{\psi}=0$. From the mathematical point of
view, the Campbell-Magaard
theorem\cite{campbell,campbellb,campbellc,campbelld} serves as a
ladder to go between manifolds whose dimensionality differs by
one. This theorem, which is valid in any number of dimensions,
implies that every solution of the 4D Einstein equations with
arbitrary energy momentum tensor can be embedded, at least
locally, in a solution of the 5D Einstein field equations in
vacuum. Because of this, the tensor $\bar{T}_{\mu\nu}$ is induced
as a 4D manifestation of the embedding geometry.

The Einstein field equations on $\Sigma _{0}$ for the metric in
(\ref{a2}), read
\begin{eqnarray}\label{a3}
r\frac{df}{dr}-1+f&=&-\frac{8\pi G}{c^2} r^{2}\rho,\\
\label{a4} r\frac{df}{dr}-1+f&=&\frac{8\pi G}{c^4} r^{2} P.
\end{eqnarray}
The resulting equation of state is
\begin{equation}\label{a7}
P= -\rho c^2 =-\frac{3c^4}{8\pi G}\frac{1}{\psi _{0}^2},
\end{equation}
which is the equation of state for a vacuum dominated by the
cosmological constant $\Lambda_0 = 1/\psi^2_0 >0$, which has been
induced from the extra space-like dimension.

If we take a constant foliation $\psi=\psi_0\neq 0$ [to avoid a
possible divergence of $\Upsilon(\psi=\psi_0)$] on the metric
(\ref{a1}) we obtain the metric(\ref{a2}), such that the effective
4D Dirac equation is given by
\begin{eqnarray}
\gamma^0 \frac{1}{c \sqrt{f(r)}} \frac{\partial
\bar{\Psi}}{\partial t} &+& \gamma^r \frac{[f(r)]^{1/4}}{r}
\frac{\partial}{\partial r} \left[ r [f(r)]^{1/4}
\bar{\Psi}\right] - \frac{\gamma^r}{r} \left( \vec{\Sigma} .
\vec{L} + \mathbb{I}\right) \bar{\Psi} + \gamma^4  m \bar{\Psi}=0,
\label{dirac4d}
\end{eqnarray}
where $\bar{\Psi}\equiv \bar{\Psi}(t,r,\theta\phi)$ and
$m<m_0/\psi_0$ should be real ($m_0 = 2,1,0,-1,-2, ...$), for any
$\psi_0\neq 0$. As a consequence of this result we obtained a
quantized mass $m$. This is an important result that shows how the
mass $m$ of the neutrinos, which are close of a background 4D SdS
metric, can be induced from a free massless 5D test spinors close
to a Ricci-flat metric in 5D. Finally, numerical solutions for
this equation were obtained in \cite{rusini}.

\section{Final Comments}

We have studied a formalism to describe 4D massive neutrinos which
are near a SdS black hole, from a 5D Ricci-flat extended SdS
metric. On this metric we define a vacuum and hence the spinors
are considered as test massless non-interacting fermion fields.
Physically, the background metric here employed describes a 5D
extension of an usual SdS static spacetime. The energy-momentum
tensor $\bar{T}_{\mu\nu}$ can be induced in 4D as a manifestation
of the embedded geometry, once we use the Campbell-Magaard
theorem. This theorem is valid in any number of dimensions and
implies that every solution of the 4D Einstein equations with
arbitrary energy momentum tensor can be embedded, at least
locally, in a solution of the 5D Einstein field equations in
vacuum (i.e., at least Ricci-flat).

Of course, this result could be extended to many other
applications, as the study of primordial neutrinos in the early
universe. This issue will be the subject of future research.
Moreover, according to the recent experimental data\cite{opera},
the study of some possible consequences for the existence of
overlighting neutrinos in the inflationary universe deserves a
careful study.

\section*{Acknowledgements}

\noindent The authors acknowledge UNMdP and CONICET Argentina for
financial support.

\end{document}